\documentstyle[epsf,floats,aps]{revtex}
\begin{document}
\twocolumn[\hsize\textwidth\columnwidth\hsize\csname
@twocolumnfalse\endcsname
\title{Mechanical properties and formation mechanisms of a wire of single gold atoms}

\author{G. Rubio-Bollinger*, S. R. Bahn$\dagger$ , N. Agra\" \i t*, K. W. Jacobsen$\dagger$ and S. Vieira*}

\address{*Laboratorio de Bajas Temperaturas, Dept. F\'\i sica de la Materia Condensada C-III,
 Instituto Universitario de Ciencia de Materiales "Nicol\' as Cabrera",
 Universidad Aut\' onoma de Madrid, E-28049 Madrid, Spain.\\
$\dagger$Center for Atomic-scale Materials Physics, Department of
Physics, Technical University of Denmark, DK-2800 Lyngby,
Denmark.}
\date{\today}
\maketitle

\begin{abstract}

A scanning tunneling microscope (STM) supplemented with a force
sensor is used to study the mechanical properties of a novel
metallic nanostructure: a freely suspended chain of single gold
atoms. We find that the bond strength of the nanowire is about
twice that of a bulk metallic bond. We perform ab initio
calculations of the force at chain fracture and compare
quantitatively with experimental measurements. The observed
mechanical failure and nanoelastic processes involved during
atomic wire fabrication are investigated using molecular dynamics
(MD) simulations, and we find that the total effective stiffness
of the nanostructure is strongly affected by the detailed local
atomic arrangement at the chain bases.

\end{abstract}


\vskip2pc]

\narrowtext

Understanding the mechanical properties of nanostructures is
essential for the atomic-scale manipulation and modification of
materials, which behave qualitatively different at the nanoscale
than at larger dimensions.  Mechanical properties of atomic-sized
metallic contacts between surface asperities have been studied
experimentally using scanning tunneling microscopy and related
techniques \cite{rubio,stalder} and theoretically in molecular
dynamics simulations \cite{landman,todorov,sorensen,stafford}.
It has been recently shown
that it is possible to extract from a previously fabricated
nanocontact stable wires of single gold atoms\cite{yanson,takayanagi},
in some cases
up to
seven atoms in length, which are freely suspended between two gold
electrodes. The conductance of these one-dimensional conductors is
close to one quantum unit of conductance $G_0=2e^2/h$ (where $e$
is the charge on an electron and $h$ is Planck's constant) because
electron transport proceeds through one single quantum conductance
channel which is almost completely open\cite{scheer}.
A very recent experimental study of gold nanocontacts at room temperature
using transmission electron microscopy shows that nanowires with a last
conductance plateau close to $G_0$ are one atom thick\cite{rodrigues}.

We have developed a specific STM supplemented with a force sensor
in order to measure the mechanical properties of these
nanostructures. The inset in Fig.~1a gives a schematic idea of the
experimental setup. The suspended gold nanowire is fabricated
between an STM gold tip and a cantilever (cylindrical gold wire,
0.125 mm diameter, 2 mm length, $>99.99 \%$ purity) and its
conductance measured applying a fixed bias voltage of 10 mV. The
experiment is performed at 4.2 K. The force during fabrication and
breaking of the wire is obtained measuring the cantilever
displacement using an auxiliary STM, working in constant tunneling
current mode, which follows the force induced displacement of the
cantilever free end with picometer resolution. In order to avoid
mechanical instabilities related to the softness of the force
sensor\cite{sorensen} and to have a rigid setup we have used a cantilever whose
spring constant is very large (380 N/m) compared to the expected
nanostructure effective elastic constant which is two orders of
magnitude smaller.

\begin{figure}[h]
\vspace{0mm}
\begin{center}
        \leavevmode
        \epsfxsize=80mm
        \epsfbox{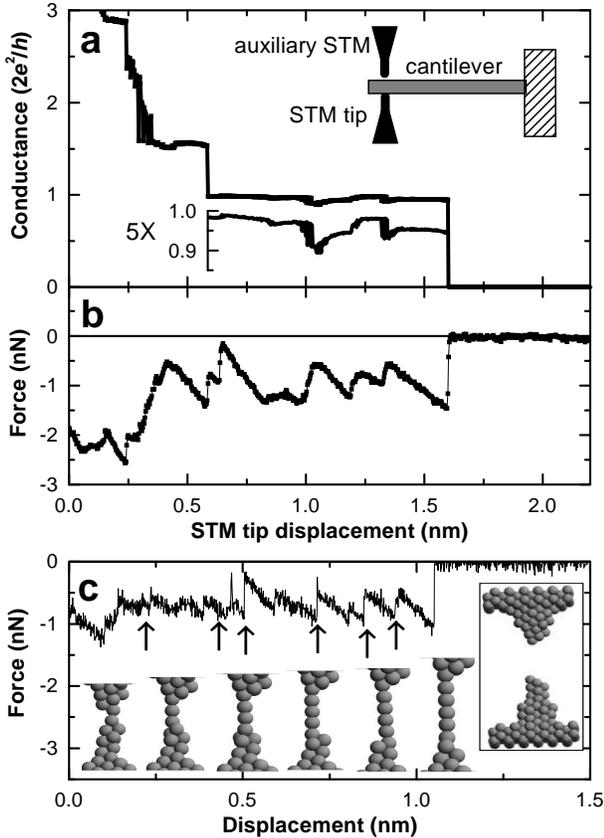}
 \end{center}
\vspace{0mm} \caption[]
 { Simultaneous conductance (a) and force (b) measurement during chain
 fabrication and breaking. The conductance in the last plateau has been
 zoomed to show detailed variations. Inset: schematic drawing of the experimental setup.
 (c) Calculated force during the MD simulation. Arrows indicate where a new atom pops
 into the chain and snapshots of the structure at these positions are shown.
} \label{FIG. 1.}
\end{figure}

An example of simultaneous measurement of conductance and force
during fabrication of a gold nanowire is shown in Fig.~1. The
experiment starts from a previously formed nanoconstriction which
is then stretched at a constant speed (0.5 nm/s). The conductance
curve displays a steplike behavior down to a value close to one
conductance quantum $G_0$, which corresponds to a one-atom contact
of gold\cite{agrait}, and the simultaneously recorded force curve shows a
sawtooth-like signal decreasing in amplitude in a sequence of
elastic stages separated by sudden force relaxations related to
atomic rearrangements in the nanocontact. This mechanical behavior
was predicted in early molecular dynamics simulations
\cite{landman} and the correspondence between sharp conductance
jumps and sudden force relaxations was observed experimentally at
room temperature down to contacts consisting of one single atom
\cite{rubio}. The connection between conductance and force has
also been addressed in several theoretical studies
\cite{todorov,sorensen,stafford,nakamura}. In the experiment shown in Fig.~1,
the one-atom contact of gold is further stretched a distance of
about 1 nm while the conductance remains close to $2e^2/h$ , which
signals the formation of a chain of about four atoms that finally
breaks \cite{yanson}. However, in contrast to the behavior
observed in metallic nanoconstrictions, the conductance remains
almost constant while the force shows large irreversible
relaxations between linear stages. Note that there are small
conductance jumps related to force relaxations, but their
magnitude is much smaller than $2e^2/h$.

\begin{figure}[h]
\vspace{0mm}
\begin{center}
        \leavevmode
        \epsfxsize=80mm
        \epsfbox{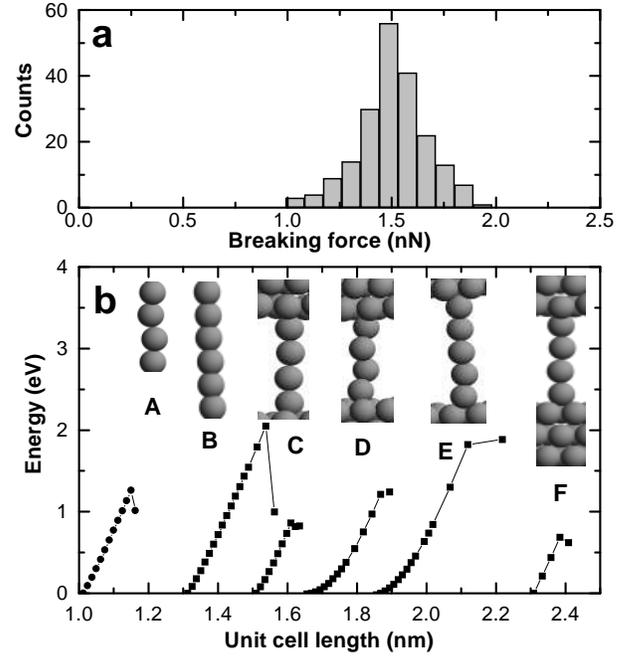}
 \end{center}
\vspace{0mm} \caption[] {(a) Histogram of force needed to break
the atomic chain, from a set of 200 chain preparation experiments.
Note also that only atomic chains whose conductance is close to Go are
considered in the histogram.
(b) The energy close to breaking for various structures as
calculated within Density Functional Theory. The resulting force
at the break point is between 1.55 nN and 1.68 nN. The various
supercells used in the calculation are shown above each curve. The
curves (B) and (E) have been shifted 0.2 units to the left and
right respectively for clarity.} \label{FIG. 2.}
\end{figure}

In order to gain some insight into the observed mechanical
behavior during the formation of an atomic chain we have performed
molecular dynamics simulations using Effective Medium Theory
(EMT)\cite{puska,stoltze}
to describe the interatomic interactions. The simulation starts
from a nanocontact cut out of an fcc crystal with a cross section
at the narrowest region corresponding to 5-10 atomic cross
sections.
Keeping the temperature at 4K the nanocontact is stretched at a rate
of 2 m/s. The simulated strain rate is thus much higher than the
experimental one leaving less time for thermally activated processes to
occur.
Upon stretching, a chain of single
gold atoms is in some cases formed by extraction of atoms from the
neighbouring electrodes into the chain. Snapshots from one such
example are shown in Fig.~1c together with the force curve
obtained from the simulation. The force curve exhibits a sawtooth
shape during the chain formation as seen in the experiments. In
the stages with a linearly growing tensile force the nanocontact
is elastically stretched while at the force jumps abrupt atomic
rearrangements occur. Some of the force jumps (generally speaking
the larger ones) correspond to incorporation of an extra atom into
the bridging atomic chain while other relaxations come from atomic
rearrangements occuring in the electrode region close to the
chain. As long as the force required to rearrange the atoms in the
electrodes is smaller than the breaking force, the chain can grow
in length. Note that both in the experimental and theoretical
curves mechanical relaxations take place at force values smaller
than the final breaking force.

\begin{figure}[h]
\vspace{0mm}
\begin{center}
        \leavevmode
        \epsfxsize=80mm
        \epsfbox{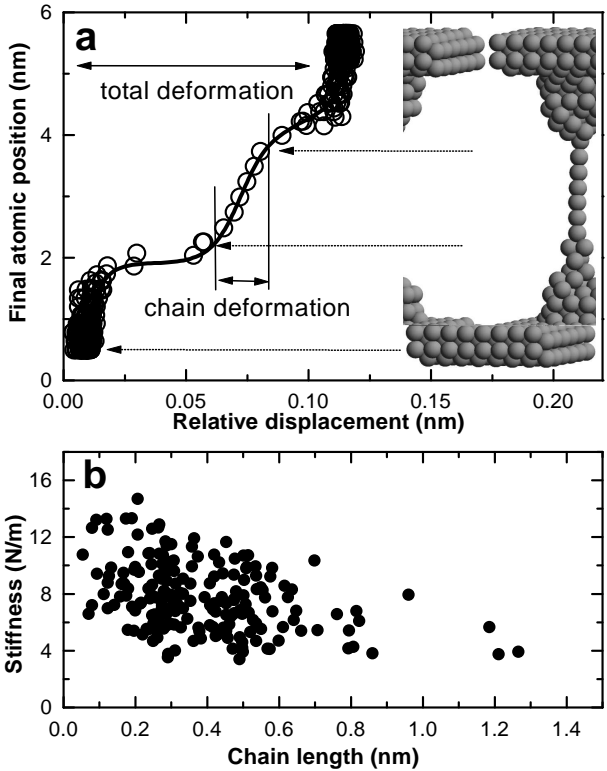}
 \end{center}
\vspace{0mm} \caption[] {(a) The relative displacement of each
atom in the structure (abscisa) during an elastic deformation
stage in the MD simulation. The ordinate gives the position of
each atom along the chain axis. To the right is shown the
configuration halfway through the stage. (b) Experimental values
of the stiffness of the nanostructure just before rupture as a
function of chain length.} \label{FIG. 3.}
\end{figure}

We have experimentally observed that the force at which the atomic chain
breaks exhibits a very stable value over two hundred nanowire
breaking experiments similar to the one shown in Fig.~1, and is
independent of chain length.
All these nanowires have a conductance close to $G_0$, and
therefore correspond to one-atom thick wires.
We obtain a narrow distribution for
the breaking force value centered at 1.5 nN. We estimate an
accuracy in the force measurement of about 20\% so this leads
to a value of 1.5 $\pm$ 0.3 nN for the force needed to break one
single bond in the chain. This allows for a
stringent test of our knowledge of the metallic bond strength in
such low dimensional systems.

In order to carry out a quantitative comparison between theory and
experiment we have performed Density Functional Theory (DFT)\cite{kohn}
calculations on small idealized contacts. The DFT calculations
have been performed on periodically repeated supercells of
different gold chain structures as depicted in Fig.~2b. Electronic
exchange and correlation effects are described within the
generalized gradient approximation\cite{perdew} and complete structural
relaxation is performed for each configuration studied\cite{campos,vanderbilt}.

The results are shown in Fig.~2b where the energy as a function of
length of the wire is plotted. We find the breaking force as the
slope of the last segment before breaking. The values found are
(in nN) (A) 1.62, (B) 1.65, (C) 1.65, (D) 1.55, (E) 1.68 and (F)
1.58. We thus find a value close to 1.6 nN regardless of the
connection of chain to a substrate slab. The lowest value (1.55
nN) is obtained for (D) while the highest (1.68 nN) is seen in the
very similar structure of (E). From this we conclude that the
influence from the slabs is marginal, in good agreement with the
narrow distribution found in the experiment. The value of 1.6 nN
is very close to the experimentally observed average value of 1.5
nN. However, it should be mentioned that the exact value depends
on the approximation for the exchange-correlation energy. Using the
RPBE approximation\cite{rpbe} we for example estimate a breaking
force of 1.4 nN.
Similar calculations have been used by other authors to
investigate the stability and morphology of the
chains\cite{torres,portal,okamoto,hakkinen,demaria}. We note
that the observed breaking force (1.5 nN) is considerably higher
than what would be expected for individual bonds in bulk gold. An
estimate of the maximal force per bond from bulk modulus and
cohesive energy considerations results in values of only 0.8-0.9
nN\cite{rose}. This is confirmed by direct density functional
equation-of-state calculations where we find a maximal force per
bulk bond of 0.71 nN. Thus the experiments give direct evidence
that bonds of low-coordinated metal atoms are considerably
stronger than bonds in the bulk. The EMT-simulations give an
average value for the force at breaking somewhat smaller (around
1.0 nN) than what is seen experimentally and in the DFT
calculations. This quantitative discrepancy is not surprising
since the EMT potential does not include shell
effects\cite{puska,stoltze}.

In order to understand the nanoelastic properties of the
chain/electrode system we have analyzed the elastic stages in the
molecular dynamics simulations in more detail. In Fig.~3a we show
the relative displacement of the atoms both in the chain and the
nearby electrodes during one of the elastic stages. As can be seen
from the figure the largest atomic displacements take place in the
electrodes, not in the chain itself.
This is due to a combination of two effects. One is the electronic
effect described above, which makes the bonds in the chain stronger
than the bonds in the more bulk like electrodes. The other effect
is geometrical. The atoms in the electrodes
are sitting in arrangements where the breaking of the bonds can
proceed not through direct radial stretching but rather through
more concerted motion of the atoms, giving longer paths and hence
smaller forces.
This also plays a role for the fact that the electrode is much softer
than the bulk.
 The consequence is the peculiar feature of the
microscopic system that thinner is actually stronger in contrast
to what is observed in macroscopic systems.
For the particular structure
shown the effective elastic stiffness of the chain is in fact 5
times larger than that of the electrodes.

The experimental indication of this point comes from studying the
slope of the force curves at the elastic stages. Fig.~3b shows a
plot of the slope of the force curve in the last elastic
deformation stage, just before chain rupture, as a function of the
length of the chain for a set of two hundred nanocontact
stretching experiments similar to that shown in Fig.~1. The data
for short chains are scattered because the elastic stage slope in
the force curve is mainly determined by the compliance of the
neighboring electrodes, and this is quite variable depending on
the exact atomic configuration of the atoms in the electrodes near
the chain ends. However, for longer chains there is less
scattering because the elastic constant of the chain becomes
comparable to that of the electrodes. Consequently the strain in
the nanowire itself cannot be directly related to relative STM tip
displacement. In particular, for short chains tip displacement
results mostly in force biasing.

Atomic chains of gold are excepcionally simple because they have a
single conductance channel which is almost completely open
irrespective of their length. Atomic wires of other metals could
have different electrical and mechanical properties due to their
more complex electronic structure, and detection of chain
formation could be more difficult because their conductance might
not be necessarily close to 1 $G_0$. The observation that low
coordinated atoms have stronger bonds than atoms in the bulk is
quite general, but the exact magnitude of this effect could have a
dramatic impact on the possibility of formation of chains in other
materials. Further experiments and calculations will be necessary
to elucidate these points.

G. R. B., N. A. and S. V. were supported by CICyT. CAMP is
sponsored by the Danish National Research Foundation.

\end{document}